\documentclass[aps,prl,superscriptaddress,preprintnumbers,reprint,nofootinbib,showpacs]{revtex4-1}
\usepackage{graphicx}
\usepackage{epsfig}
\usepackage{amssymb}

\def\als{\alpha_{\rm s}} 
\def\MS{\overline{\rm MS}}
\newcommand{\be}{\begin{equation}}
\newcommand{\ee}{\end{equation}}
\newcommand{\bea}{\begin{eqnarray}}
\newcommand{\eea}{\end{eqnarray}}

\begin{document}

\title{Precision determination of $r_0\Lambda_{\MS}$ from the QCD static energy}
\author{Nora Brambilla}
\affiliation{Physik Department, Technische Universit\"at M\"unchen, D-85748 Garching, Germany}
\author{Xavier \surname{Garcia i Tormo}}
\affiliation{Department of Physics, University of Alberta, Edmonton,
  Alberta, Canada T6G 2G7} \altaffiliation[Current address: ]{Institut f\"ur Theoretische Physik, Universit\"at Bern,
  Sidlerstrasse 5, CH-3012 Bern, Switzerland.}
\author{Joan Soto}
\affiliation{Departament d'Estructura i Constituents de la Mat\`eria and Institut de Ci\`encies del Cosmos, 
Universitat de Barcelona, Diagonal 647, E-08028 Barcelona, Catalonia, Spain}
\author{Antonio Vairo}
\affiliation{Physik Department, Technische Universit\"at M\"unchen, D-85748 Garching, Germany}

\date{\today}

\preprint{Alberta Thy 04-10,~   TUM-EFT 8/10,~   UB-ECM-PF 10/15,~  ICCUB-10-029}

\begin{abstract}
We use the recently obtained theoretical expression for the complete
QCD static energy at next-to-next-to-next-to leading-logarithmic
accuracy to determine $r_0\Lambda_{\MS}$ by comparison with available
lattice data, where $r_0$ is the lattice scale and $\Lambda_{\MS}$ is the QCD scale. 
We obtain $r_0\Lambda_{\MS}=0.637^{+0.032}_{-0.030}$ for the zero-flavor case. 
The procedure we describe can be directly used to obtain $r_0\Lambda_{\MS}$ 
in the unquenched case, when unquenched lattice data for the static energy 
at short distances becomes available. 
Using the value of the strong coupling $\als$ as an input, the unquenched result would
provide a determination of the lattice scale $r_0$.
\end{abstract}


\maketitle

The energy between a static quark and a static antiquark is a
fundamental object to understand the behavior of quantum chromodynamics
(QCD) \cite{Wilson:1974sk}. Its long-distance part encodes the confining dynamics of the
theory while the short-distance part can be calculated to high
accuracy using perturbative techniques. Perturbative computations of
the short-distance part have been performed for many years
\cite{Fischler:1977yf,Billoire:1979ih} and the
two-loop corrections have been known for quite some time now \cite{Peter:1996ig,Peter:1997me,Schroder:1998vy}.
When using perturbation theory to calculate the short-distance part, the virtual emission of gluons
that can change the color state of the quark-antiquark pair (so-called ultrasoft gluons) 
produce infrared divergences, which induce logarithmic terms, $\ln\als(1/r)$, in the static energy. 
Those effects, which first appear at the three-loop order, were identified in
Ref.~\cite{Appelquist:1977es} and calculated in Ref.~\cite{Brambilla:1999qa,Kniehl:1999ud} 
using an effective field theory framework \cite{Pineda:1997bj,Brambilla:1999xf}. 
That framework also allows for resummation of the
ultrasoft logarithms \cite{Pineda:2000gza}, which may be large at small
distances $r$. Very recently, the complete three-loop corrections to the
static energy have become available \cite{Smirnov:2008pn,Anzai:2009tm,Smirnov:2009fh}. 
Combining the results of those calculations with the resummation of the ultrasoft
logarithms at sub-leading order \cite{Brambilla:2009bi,Brambilla:2006wp}, the static energy at
next-to-next-to-next-to leading-logarithmic (N$^3$LL) accuracy,
i.e. including terms up to order $\als^{4+n}\ln^n\als$ with $n\ge 0$, is now
completely known. 

In the first part of the letter, we compare 
the static energy at N$^3$LL accuracy with lattice data.
The comparison shows that, after subtracting the leading
renormalon singularity, perturbation theory 
 reproduces very accurately 
the lattice data at short distances, thus confirming at an unprecedented precision level the
 conclusions reached in previous analyses \cite{Brambilla:2009bi,Pineda:2002se,Sumino:2005cq}. 
In the second part of the letter, the excellent agreement of perturbation theory with lattice data
 allows us to obtain a precise determination of the quantity $r_0\Lambda_{\MS}$, where $r_0$ is the
lattice scale and $\Lambda_{\MS}$ is the QCD scale (in the $\MS$
scheme), a key ingredient to relate low energy hadronic physics with high energy collider phenomenology. This constitutes the main result of our work.

The static energy $E_0(r)$ at short distances can be written as
\begin{equation}\label{eq:E0}
E_0(r)=V_s+\Lambda_s+\delta_{\text{US}},
\end{equation}
where $V_s$ and $\Lambda_s$ are matching coefficients in potential
Non-Relativistic QCD (pNRQCD) \cite{Brambilla:1999xf} and $\delta_{\text{US}}$ contains the
contributions from ultrasoft gluons. $V_s$ corresponds to the static
potential and $\Lambda_s$ inherits the residual mass term from the
Heavy Quark Effective Theory Lagrangian. In order to obtain a rapidly converging
 perturbative series for the static potential in the short-distance
regime, it has been argued that it is necessary to implement a scheme
that cancels the leading renormalon singularity \cite{Beneke:1998rk}. The use of any such scheme
introduces an additional dimensional scale (which we call $\rho$), upon
which all the quantities in Eq.~(\ref{eq:E0}) depend. We will employ the
so-called RS scheme \cite{Pineda:2001zq}, in the same way as it was done in
Ref.~\cite{Brambilla:2009bi}. The explicit expressions for $E_0$ at
N$^3$LL accuracy were presented in Ref.~\cite{Brambilla:2009bi} and will not be repeated here. We
refer to that paper for details. The only new ingredient is that the
three-loop gluonic contribution to the static potential is now known. At
three-loop order the static potential presents infrared divergences,
which cancel in the physical observable $E_0$ after the inclusion of
the ultrasoft effects. Therefore, it is necessary to consistently use the same
scheme to factorize the ultrasoft contributions for all the terms in
Eq.~(\ref{eq:E0}). That way one obtains the correct three-loop
coefficient for the static energy $E_0$, which is independent of the
scheme used to factorize the ultrasoft
contributions. Refs.~\cite{Anzai:2009tm,Smirnov:2009fh} present the
result for the purely-gluonic three-loop coefficient of the static
potential in momentum space, which we call $a_3^{(0)}$ (following the
notation of Ref.~\cite{Smirnov:2009fh}). We emphasize again that
$a_3^{(0)}$ is scheme dependent. The corresponding coefficient
in the static energy can be obtained by taking the $d$-dimensional
Fourier transform of the momentum-space potential 
(as calculated in Refs.~\cite{Anzai:2009tm,Smirnov:2009fh}) and adding to it the ultrasoft
contribution. In the factorization scheme used in
Refs.~\cite{Anzai:2009tm,Smirnov:2009fh}, the ultrasoft contribution is given by Eq.~(8) of
Ref.~\cite{Anzai:2009tm} (which we 
confirm). Note that the scheme used in
Refs.~\cite{Anzai:2009tm,Smirnov:2009fh} is different from the one
used in Ref.~\cite{Brambilla:2009bi}. If we
then subtract Eq.~(34) of Ref.~\cite{Brambilla:2009bi} (the ultrasoft
contribution in the scheme of that paper) from 
the static energy we get the three-loop gluonic contribution
to the static potential in the scheme of
Ref.~\cite{Brambilla:2009bi} (which we will denote as
$a_{3,\textrm{\tiny Ref.\cite{Brambilla:2009bi}}}^{(0)}$). By doing
that we obtain \cite{Smirnov:2009fh} $c_0:=a_{3,\textrm{\tiny
    Ref.\cite{Brambilla:2009bi}}}^{(0)}/4^3=222.703$.
This was the only missing ingredient of the static energy at N$^3$LL
accuracy at the time that paper was written. Note that the value of $c_0$ above is within the
range (215,350) predicted in Ref.~\cite{Brambilla:2009bi}, and considerably lower than the Pad\'e estimate $c_0=313$ commonly used in the literature \cite{Chishtie:2001mf}.

\begin{figure}
\centering
\includegraphics[width=8.6cm]{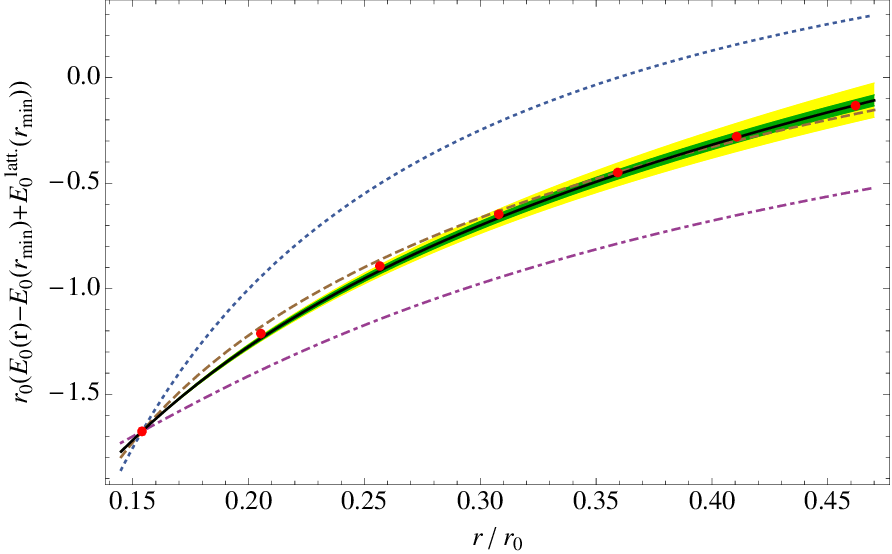}\\
(a)\\

\vspace{.5cm}

\includegraphics[width=8.6cm]{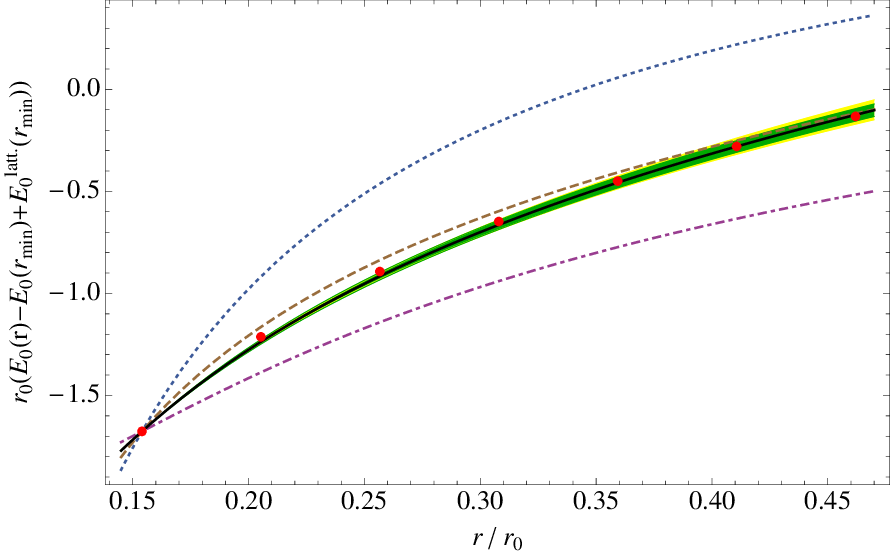}\\
(b)
\caption{(a) Comparison of the singlet static energy with lattice
  data. We plot $r_o\left(E_0(r)-E_0(r_{\rm min})+E_0^{\rm latt.}(r_{\rm min})\right)$
 as a function of $r/r_0$ and the
  lattice data of Ref.~\cite{Necco:2001xg} (red points). The dotted blue
  curve is at tree level, the dot-dashed magenta curve is at one loop, the
  dashed brown curve is at two loop plus leading ultrasoft logarithmic
  resummation and the solid black curve is at three loop plus next-to-leading 
  ultrasoft logarithmic resummation. The yellow (lighter) band is
  obtained by varying $r_0\Lambda_{\MS}$ according to
  $r_0\Lambda_{\MS}=0.602\pm0.048$ \cite{Capitani:1998mq} and the green
  (darker) band is obtained by adding the term $\pm C_F\als^5/r$, for
  the solid black curve.
  (b) Same but using $r_0\Lambda_{\MS}=0.637^{+0.032}_{-0.030}$ (see text).
}\label{fig:sten}
\end{figure}

We now compare the perturbative results for the static energy with
the $n_f=0$ lattice data of Ref.~\cite{Necco:2001xg}. All results are presented in units of $r_0$. 
As it is explained in Ref.~\cite{Brambilla:2009bi}, an appropriate quantity to
plot for this comparison is
\begin{equation}
\label{eq:Etilde}
E_0(r)-E_0(r_{\rm min})+E_0^{\rm latt.}(r_{\rm min}),
\end{equation}
where $r_{\rm min}$ is the shortest distance at which lattice data is
available and $r_0E_0^{\rm latt.}(r_{\rm min})=-1.676$
\cite{Necco:2001xg}. Now that the three-loop static potential is known
the normalization of the $u=1/2$ renormalon singularity $R_s$, which is
necessary to implement the RS scheme, can be determined using one
order more in the perturbative expansion of the potential. We obtain
\begin{equation}
R_s=-1.333+0.499-0.338-0.033=-1.205, 
\end{equation}
which is the value we will use. The rest of the scales and parameters are set as in
Ref.~\cite{Brambilla:2009bi}. We note that, in particular, this means
that we have $\rho=3.25r_0^{-1}$ and $r_0\Lambda_{\MS}=0.602$ \cite{Capitani:1998mq}. The
comparison of the static energy with lattice data is presented in
Fig.~\ref{fig:sten}(a). We recall that all the curves coincide with
the lattice point at $r=r_{\rm{min}}$ by construction [as one can see
from Eq.~(\ref{eq:Etilde})]. We note that, due to the singlet-octet mixing
in the renormalization group equation for $\Lambda_s$ (at order $r^2$
in the multipole expansion of pNRQCD), the N$^3$LL curve (solid black)
depends on a
constant, which we call $K_2$. Power counting tells us that $|K_2|\sim\Lambda_{\MS}$, but the constant is otherwise arbitrary. 
We fix it by a fit to the lattice data, which delivers
$r_0K_2=-2.06$. The bands in Fig.~\ref{fig:sten}(a)
 illustrate the effect of variations in
$r_0\Lambda_{\MS}=0.602\pm0.048$ \cite{Capitani:1998mq} (yellow lighter band) and the effect of adding the term $\pm C_F\als^5/r$,
which is representative of the neglected higher
order corrections (green darker band); the remaining parameters are
kept at their original values to obtain the bands.
The figure confirms that perturbation theory (in the RS scheme) reproduces
very well the lattice data for the static energy at short
distances. We also note that the band due to the neglected
higher order terms is smaller than the one due to the uncertainty in
$r_0\Lambda_{\MS}$. Those facts 
indicate that we should be able to use the lattice
data to obtain a more precise determination of $r_0\Lambda_{\MS}$, as
we describe later. It is worth emphasizing that the aim of Fig.~\ref{fig:sten} is to
compare with lattice data for the specific $r$ range displayed in the
figure and to see if the theoretical curves 
follow the lattice data points starting from the point at
the shortest distance. The scale $\rho$ was set to the fixed value $3.25r_0^{-1}$, which 
is at the center of the range, and it was kept the same for all curves. 
Moreover, 
the static energy was expressed as a series in
$\als(\rho)$, rather than $\als(1/r)$, to reduce the uncertainty
associated with $R_s$ and the specific implementation of the renormalon
subtraction. 

\begin{figure}
\centering
\includegraphics[width=8.6cm]{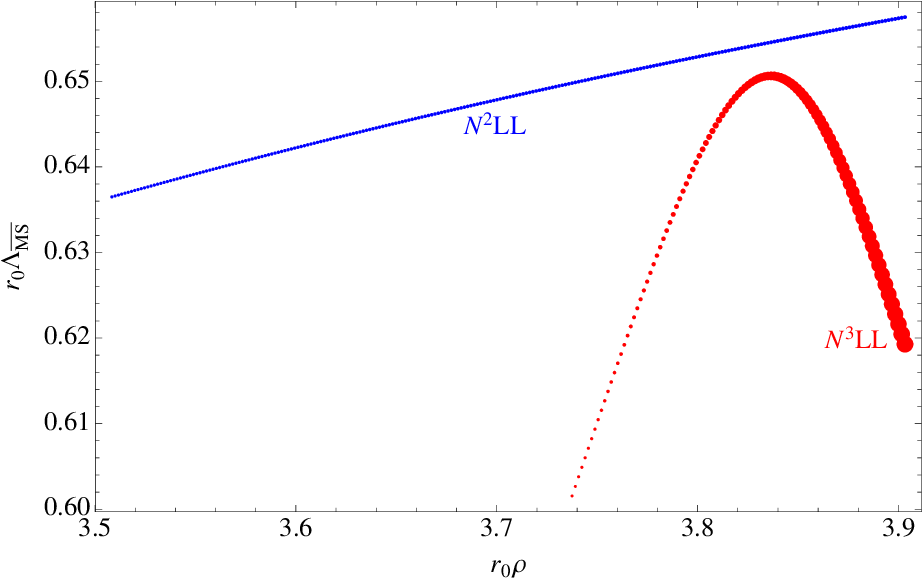}
\caption{Fit values of $r_0\Lambda_{\MS}$ for different values of
  $\rho$. The range of $\rho$ displayed in the plot corresponds to that
  obtained after step 3 (N$^2$LL points) or step 4 (N$^3$LL points) of
  the procedure described in the text. The size of each point is
  proportional to its assigned weight.
}\label{fig:rholam}
\end{figure}

We will now assume that perturbation theory by itself (after canceling
the leading renormalon) is indeed enough
to accurately describe the lattice data for the range of $r$ we are
considering, i.e. $0.15r_0 \le r<0.5r_0$. That is, we are assuming that non-perturbative 
effects are small and can be totally neglected \footnote{The leading
  genuine non-perturbative contribution, which is proportional to the
  gluon condensate, is of order $\Lambda_{\MS}^4 r^3/\als (1/r)$, and
  hence parametrically suppressed according to the counting of
  Ref.~\cite{Brambilla:2009bi}, which assumes $ \Lambda_{\MS} \sim \als^2/r$.}.
With this assumption, we can use the lattice data for the static energy to
determine $r_0\Lambda_{\MS}$. We initially consider 
the set of $r_0\Lambda_{\MS}$ for which: (i) the perturbative series
for the static energy appears to converge, and (ii) the agreement with
lattice is improved when
increasing the perturbative order of the calculation.
If we implement the condition (ii) by demanding that the reduced $\chi^2$ of the
curves decreases when we increase the perturbative order of the
calculation, we obtain the range (0.58,0.8) for
$r_0\Lambda_{\MS}$. We now proceed to improve the precision of that
determination. First we recall that the use of the RS scheme introduced the scale $\rho$ in
the expressions for the static energy. We have used the value
$\rho=3.25r_0^{-1}$ because it corresponds to the inverse of the
central value of the $r$ range we are comparing with lattice
data. That way we keep $\ln r\rho$ terms from becoming large \cite{Brambilla:2009bi}, but in
principle any $\rho$ around that value is valid. 
We will exploit this freedom to find a set of $\rho$ values which are ``optimal'' for
the determination of the parameter $r_0\Lambda_{\MS}$
by following the procedure we describe next:
\begin{enumerate}
\item We vary $\rho$ by $\pm 25\%$ around $\rho=3.25r_0^{-1}$,
  i.e. from $\rho=2.44$ to $\rho=4.06$.\footnote{We have used steps
    of $1.6\times 10^{-3}r_0^{-1}$ to do that.}
\item For each value of $\rho$ and at each order in the perturbative
  expansion of the static energy, we perform a fit to the lattice
  data. The parameters of the fit are $r_0\Lambda_{\MS}$ for the
  curves from tree level to next-to-next-to leading logarithmic (N$^2$LL)
  accuracy (1-parameter fits), and $r_0\Lambda_{\MS}$ and $r_0K_2$ for the
  curve at N$^3$LL accuracy (2-parameter fit).
\item We select those $\rho$ values for which the reduced $\chi^2$ of
  the fits decreases when increasing the order of the perturbative expansion.
\item Finally, from the set of $\rho$ values obtained above, we select
  the ones for which the fitted value of $K_2$ is compatible with the
  power counting (we require $|r_0K_2|\le 2$ \cite{Brambilla:2009bi}).
\end{enumerate}
The above steps provide us with a certain range of $\rho$. We then consider the set of fitted values of $r_0\Lambda_{\MS}$ at N$^3$LL
accuracy (denoted as $x_i$ below) for that range.
In order to give more significance to the better fits, we assign a
weight to each
of the $x_i$. We choose those weights to be given by the inverse of the
reduced $\chi^2$ of the fit.
We take the weighted average of the $x_i$ as our central value for the
determination of $r_0\Lambda_{\MS}$ and obtain
\begin{equation}
\bar{x}:=\sum_i w_i x_i=0.637,
\end{equation}
where $w_i:=\tilde{w}_i/(\sum_j\tilde{w}_j)$ and $\tilde{w}_i$ is the weight of the point $x_i$.
To estimate the error that we should associate
to that number, we first consider the weighted standard deviation of
that set of values and assign it as an error to the weighted
average\footnote{A similarly motivated procedure to estimate theoretical errors has been
  used, for example, in Ref.~\cite{Durr:2010hr}.}, we obtain
\begin{equation}
\sigma:=\sqrt{\frac{1}{1-\sum_j\!w_j^2}\sum_iw_i(x_i-\bar{x})^2}=0.012.
\end{equation}
Fig.~\ref{fig:rholam} shows the obtained fit values of
$r_0\Lambda_{\MS}$ at N$^3$LL accuracy and at N$^2$LL accuracy, for the different values
of $\rho$. The size of each point in the plot is proportional to its
weight. Additionally, we also consider the difference between the
weighted averages computed using the N$^3$LL result and the N$^2$LL
result and assign it as a second error to the weighted average
at N$^3$LL accuracy (we compute the weighted average at
N$^2$LL accuracy in the $\rho$ range obtained after step 3 of the procedure
described above, since step 4 applies only to the N$^3$LL result).  
The value of that difference is 0.011. We sum those
two errors linearly and obtain $0.637\pm0.023$. We present in Table \ref{tab:Lam} the
obtained values for $r_0\Lambda_{\MS}$ following this procedure at
different orders of accuracy.  We emphasize that the error assigned to the result must account for the uncertainties
associated to the neglected higher order terms in the perturbative
expansion of the static energy; in that sense, we note that Table
\ref{tab:Lam} shows that our errors
at a certain perturbative order always include the central value at
the next order, which gives us confidence in the reliability of the procedure. In order to further assess the systematic errors steming from our
procedure, we have
 redone the analysis using two additional weight assignements:
 (1) $p$-value weights, this analysis gives $r_0\Lambda_{\MS} =
 0.638\pm0.031$, and (2) constant weights,  this analysis gives
 $r_0\Lambda_{\MS} = 0.634\pm0.028$.
 These two numbers are compatible with our previous value. In our
final result, we will quote an error that covers the whole range
spanned by the three analyses,
 which we consider a fairly conservative estimate.
 Finally, we mention that there is also an error associated to
each of the fit values $x_i$ coming from the variance of the lattice data
points\footnote{To obtain this error we consider 1-parameter fits,
  with $K_2$ fixed.}. The error induced in this way in $\bar{x}$ is much smaller than
the other ones we are considering (due to the fact that the lattice points 
have very small error bars) and can be neglected.
\begin{table}
\begin{tabular}{c|c}
Accuracy & $r_0\Lambda_{\MS}$ \\
\hline next-to-leading & $0.946\pm0.039\pm0.80$ \\
N$^2$LL & $0.648\pm0.006\pm0.30$ \\
N$^3$LL (no p.c.c.) & $0.619\pm0.04\pm0.029$ \\
N$^3$LL & $0.637\pm0.012\pm0.011$ \\
\end{tabular}
\caption{Values of $r_0\Lambda_{\MS}$ obtained at different levels of
  accuracy. N$^3$LL (no p.c.c.) stands for N$^3$LL accuracy without
  imposing the power counting constraint on $K_2$, i.e. omitting step
  4 in the procedure described in the text. The first error corresponds
to the weighted standard deviation and the second one to the
difference with the previous order.}\label{tab:Lam}
\end{table}

According to the discussion above, our final result for $r_0\Lambda_{\MS}$ reads
\begin{equation}\label{eq:ourlam}
r_0\Lambda_{\MS}=0.637^{+0.032}_{-0.030}.
\end{equation}
Our result is compatible with the value
$r_0\Lambda_{\MS}=0.602\pm0.048$ given in Ref.~\cite{Capitani:1998mq},
which we had been using previously, but has a smaller error. We also
mention that a new (preliminary) lattice calculation, which determines 
$r_0\Lambda_{\MS}$ from the ghost and gluon propagators, 
obtains a result, $r_0\Lambda_{\MS}=0.62 \pm 0.01$, which is similar to ours \cite{Sternbeck:2010xu}. 
We present in Fig.~\ref{fig:sten}(b) a comparison of the
static energy with lattice data using the value obtained in
Eq.~(\ref{eq:ourlam}) for $r_0\Lambda_{\MS}$, and the best fit value 
for $r_0K_2$ when that value of $r_0\Lambda_{\MS}$ is used, namely
$r_0K_2=-1.39$. We note that the error bands due to uncertainties in $r_0\Lambda_{\MS}$ and higher-order terms in Fig.~\ref{fig:sten}(b)
are of comparable size.

We would like to emphasize that exactly the same procedure we have described here could
be used in the unquenched case. If unquenched lattice data for the
static energy at short distances were available we could obtain an unquenched value for
$r_0\Lambda_{\MS}$ from it. Combining that result with the value of
the strong coupling $\als$, determined from other sources, would
provide a determination of the lattice scale $r_0$. This determination would be model independent and alternative to other determinations like, for instance, the one using the 1S-2S bottomonium radial excitation energy \cite{Davies:2009tsa}.

In summary, we have updated the comparison of the static energy at N$^3$LL
accuracy with lattice data by including the recently calculated
purely-gluonic  three-loop contribution to the static potential, which
was the last missing ingredient, (see
Fig.~\ref{fig:sten}). We confirmed that, after canceling the leading
renormalon singularity, perturbation theory can accurately 
reproduce the lattice data at short distances. 
By taking advantage of this fact, we have obtained a
new determination of $r_0\Lambda_{\MS}$. 
Our result for the zero-flavor case is $r_0\Lambda_{\MS}=0.637^{+0.032}_{-0.030}$, which improves the precision of the N$^2$LL determination (see table \ref{tab:Lam}) by
an order of magnitude.

\begin{acknowledgments}
We thank Federico Mescia and Alberto Ramos for useful discussions. The research of X.G.T.\ was supported by Science
and Engineering Research Canada. N.B., J.S. and A.V. acknowledge
financial support from the RTN Flavianet MRTN-CT-2006-035482
(EU). N.B. and A.V. acknowledge
financial support from the DFG cluster of excellence ``Origin and structure of the universe''
(\href{http://www.universe-cluster.de}{www.universe-cluster.de}). J.S. also acknowledges financial 
support from the ECRI HadronPhysics2 (Grant Agreement
n. 227431) (EU), the  FPA2007-60275/ and FPA2007-66665-C02-01/
MEC grants, the Consolider Ingenio program CPAN CSD2007-00042 (Spain), and the 2009SGR502 CUR grant (Catalonia).
\end{acknowledgments}

\end{document}